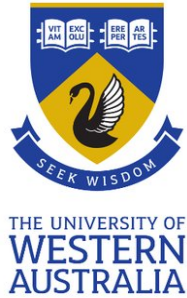

# Automated Simulations of Galaxy Morphology Evolution using Deep Learning and Particle Swarm Optimisation

Eleanor Leung (21149831)
*School of Computer Science and Software Engineering*
*The University of Western Australia*

Dr. Kenji Bekki
*ICRAR M468*
*The University of Western Australia*

Dr. Lyndon While
*School of Computer Science and Software Engineering*
*The University of Western Australia*

Word Count: 7881

# ABSTRACT

The formation of Hoag-type galaxies with central spheroidal galaxies and outer stellar rings has yet to be understood in astronomy. We consider that these unique objects were formed from the past interaction between elliptical galaxies and gas-rich dwarf galaxies. Dr. Kenji Bekki has modelled this potential formation process through simulation. These numerical simulations are a means of investigating this formation hypothesis, however the parameter space to be explored for these simulations is vast. Through the application of machine learning and computational science, we implement a new two-fold method to find the best model parameters for stellar rings in the simulations. First, test particle simulations are run to find a possible range of parameters for which stellar rings can be formed around elliptical galaxies (i.e. Hoag-type galaxies). A novel combination of particle swarm optimisation and Siamese neural networks has been implemented to perform the search over the parameter space and test the level of consistency between observations and simulations for numerous models. Upon the success of this initial step, we subsequently run full chemodynamical simulations for the derived range of model parameters in order to verify the output of the test particle simulations. We successfully find parameter sets at which stellar rings can be formed from the interaction between a gas-rich dwarf galaxy and a central elliptical galaxy. This is evidence that supports Dr. Bekki's hypothesis about the formation process of Hoag-type galaxies. In addition, this suggests that our new two-fold method has been successfully implemented in this problem search-space and can be investigated further in future applications.

# TABLE OF CONTENTS

# 1    INTRODUCTION

## 1.1    Hoag's Object

Galaxy morphology refers to the structural properties of galaxies, and is the basis of a classification system used by astronomers to group galaxies by visual appearance. By systematically examining the finer details of galaxy morphology, it is possible to gain a deeper understanding of how galaxies evolve and how galaxy types may be related to one another. Among these details are ring features, believed to have the potential to reveal information about galaxy evolution and dynamics [1].

Hoag's Object is an atypical instance of a ring galaxy where the outside ring consists of young blue stars that circle the older yellow nucleus of the galaxy. Between the two is a gap almost completely void of stars. In 1950, Astronomer Art Hoag identified Hoag's Object 600 million light-years away in the constellation Serpens [2]. Since the discovery of this phenomenon, researchers have speculated on how this galaxy could have formed - some have hypothesised a galaxy collision billions of years ago, while others have proposed that Hoag's Object is the product of extreme "bar instability" that occurred billions of years ago in a barred spiral galaxy [3]. However, these theories have been disproved or concluded to be an unlikely hypothesis.

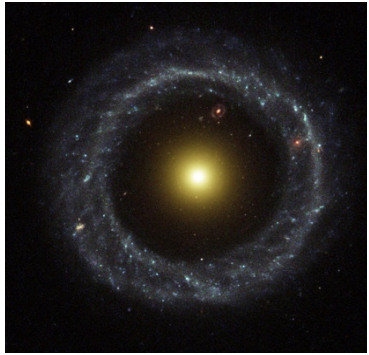

**Figure 1**: Hoag's Object as taken from NASA's Hubble Space Telescope [5].

Due to the mystery surrounding Hoag's Object and its formation, attempts in astronomy to study similar ring-type galaxies have been published. There are a few galaxies which have been identified and share fundamental characteristics with Hoag's Object, such as having a bright detached ring of blue stars. However, the centres of these galaxies have been found to exhibit different properties, or do not align with the symmetry of Hoag's Object. This group of galaxies is referred to as *Hoag-type galaxies*, and is extremely rare [4].

## 1.2    Deep Learning

Deep learning has recently attracted much attention due to a surge of development in the field's research, literature and application. Despite artificial neural networks and the backpropagation algorithm –the foundations upon which deep learning is built upon – having been used as early as the 1940's [6] and 1960's [7] respectively, it is modern day big data and computation power that has popularised the field. Originally inspired by biological neural networks in the brain, neural networks are mathematical models that are able to recognise complex relationships between inputs and outputs. They are often referred to as universal function approximators and have successfully become the state-of-the-art for several kinds of tasks spread across multiple industries [8, 9]. With the influx of research and development, different types of neural network architectures have branched out and have found much success. Convolutional neural networks are an example of this; they are used in image based tasks that involve classification or detection. Siamese networks are another variation of neural network architecture that have been used in tasks such as facial recognition and signature verification. This research paper will utilise both convolutional neural networks and Siamese networks.

## 1.3    Search Algorithms

Search algorithms and optimisation techniques are fundamental tools in classical mathematics, science and computer science problems [10]. On one hand, new and novel search algorithms are constantly created for specific use cases and applications. In contrast, established techniques continue to provide valuable results [29]. When a problem's parameter search-space is continuous, it should follow that the output is also continuous if the process from input to output is purely numerical. Under these circumstances, it is possible to use a subset of algorithms known as gradient-based algorithms which utilise calculus as a foundation for iteratively improving the outcome of the result.

# 2    KEY TERMS

**Table 1:** A list of key terms and their definitions.

| Term | Definition |
|------|------------|
| Neural Network | A neural network is a statistical data modelling tool that can model complex relationships between inputs and outputs. Neural networks are composed of elements known as 'neurons' which apply an activation function to an input. Neurons are organised into layers – an input layer, hidden layers and an output layer. |
| Loss Function | A function that measures the inconsistency between the predicted output and the actual output. The loss function optimises the parameter values in a neural network. |
| Activation Function | A bounded function that is applied to the weighted and summed inputs and transforms the output signal to reduce its range and introduce non-linearity. |
| Convolutional 2d Layer | A convolutional layer consists of a set of 2-dimensional filters or kernels that are convolved across an input image to generate a 2-dimensional activation map of the filter. A neural network learns filters that identify specific features of an image. |
| Max-Pooling Layer | A max-pooling layer reduces the dimensionality of a feature space, retaining only the most important information. The maximum value is selected from a region of pixels. |
| Dropout Layer | A dropout layer is a regularisation technique that combats overfitting by randomly setting a fraction of neurons to 0 at every training iteration. |
| Fully Connected Layer | A neural network layer where neurons are connected to all activations in the previous layer. |
| Hoag-type galaxy | A class of galaxies typically characterised by a core surrounded by a nearly perfect ring of blue stars, named after Hoag's Object. |

# 3    PROJECT SCOPE

Astronomers and astrophysicists have relied on large scale simulations and computer programs to create models of how the universe works since the 1980s [11]. Researchers currently do not yet have the insight to explain certain phenomena - such as how certain types of galaxies are formed. Hoag-type galaxies are an example of where we lack such understanding.

Dr. Kenji Bekki proposes that the formation of Hoag-type galaxies occurs from the dynamical interaction between elliptical galaxies and gas-rich dwarf galaxies, and has written computer simulations that model this interaction. The outcome of the simulations is dependent on a number of parameters that determine the interaction process. The simulations produce 3-dimensional (3d) position and velocity data for each particle, however this information is impossible to correctly determine for galaxies observed from Earth as we can only deduce the quantity and Doppler shift of light from the stellar body. As such, it is not possible to take an observed galaxy and reverse the simulation to model how it formed. Instead it is necessary to search through all possible parameters that could lead to the formation of a Hoag-type galaxy.

By automating the process of exploring the broad parameter space of the simulations, time is saved from analysing results of the simulation one at a time and manually making decisions on how to manipulate the parameter space. Therefore, the aim of this work is to determine the parameter set – the initial 3d position and 3d velocity of the dwarf galaxy – for Dr. Bekki's simulations that would most closely result in an output resembling a Hoag-type galaxy using an automated search algorithm. This can be conceptualised as an optimisation problem, where the goal is to find the global optimum amongst the parameter space. Figure 2 illustrates this entire process from an end-to-end perspective. The scope of this project is encapsulated by steps 5 and 6.

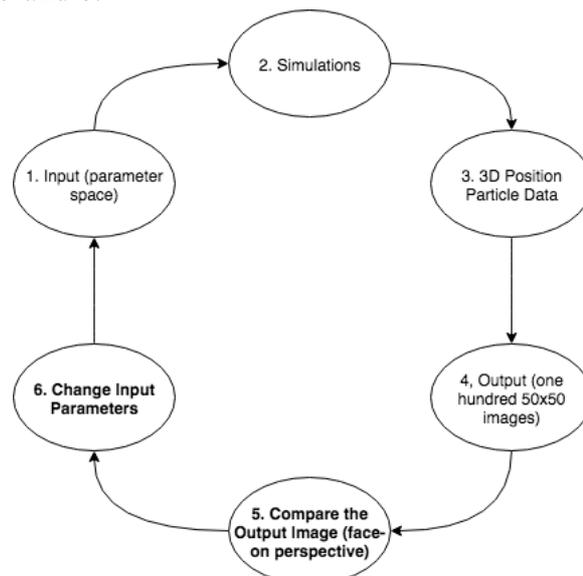

**Figure 2**: An end-to-end representation of the scope of the project problem. Steps 1 – 4 are provided by Dr. Bekki and steps 5 – 6 are to completed within this research.

Upon the success of this first goal, the final step involves taking the best set of parameters and validating the results. This is done by passing the parameter set into more complex and accurate simulations that also model the same interaction, but additionally account for gravitational dynamics *and* hydrodynamics. The output of this more complex, accurate but time-consuming simulation will be expected to be a close representation of a Hoag-type galaxy.

# 4    LITERATURE REVIEW

## 4.1    Astronomy

**Observational Studies of Ring Galaxies**

Rings are features of galaxies that have the potential to provide a greater insight to galaxy evolution and dynamics [1]. Galactic rings have been observed to have at least 3 basic types - nuclear, inner and outer rings. Current research on ring galaxies range from the classification of ringed galaxies (Buta 2017), to photometric analysis on observed ringed galaxies [4]. Early studies of this phenomena look into their physical properties, dynamics and methods of formation [12, 13]. For example, in more recent research, a study carried out by Buta in 2017 known as the Galaxy Zoo Project managed to classify 3962 ringed galaxies from the Galaxy Zoo 2 database in order to further the understanding of nature and secular evolution of ringed galaxies.

This catalogue of galaxies is not intended to be used for general statistical studies or any kind of study that required complete samples, as it is highly focussed on ringed galaxies. However, the value in Buta's research is that it helps to identify and differentiate between 'good' examples of rings and extreme examples of rings. It is from this catalogue that a ring galaxy similar to Hoag's Object was found; a large, blue ring enveloping an elliptical galaxy.

The philosophy of the project driving Buta's work is extremely interesting with respect to the proposed thesis, as it suggests that human visual classification is superior to computer classification at reliably identifying major and minor classes of galaxies. Humans may also recognise special cases of interest (like Hoag's Object) that a computer program would likely overlook [1]. This thesis is attempting to challenge this notion.

**Hoag's Object**

The mysterious nature of Hoag's Object has resulted in research into its morphology and formation process. O'Connell, Scargle and Sargent (1974) conducted a study on Hoag's Object that eliminated the possibility that the absorption or scattering by dust particles surrounding the galaxy are the cause for its ring-like structure [14]. Instead, they proposed that the ring was formed from the aftermath of an encounter with a companion or passing galaxy. This idea of galaxy interaction was further researched by Schweizer *et al.* (1987) – they found that this scenario could be unlikely due to the relatively low velocity between the galaxy's core and its ring [15]. Despite the multitude of studies and hypotheses, there has been no general consensus on this area of research. However, the interest in Hoag's Object became the inspiration for a class of galaxies with bright external rings which appear to be detached from a round core. Several discoveries of observed instances of this class of galaxies, known as Hoag-type galaxies, has led to numerous studies of their morphologies and properties.

**Hoag-type Galaxies**

Detailed studies of Hoag-type galaxies have been presented in literature revealing valuable information about the nature of the structural components of this galaxy type. The study of peculiar systems such as Hoag-type galaxies can improve our understanding of galaxy formation in general as they represent extreme cases and provide clues on formation mechanisms [4]. One such exemplar includes Pakdil *et al*. (2017), who conducted a photometric study of PGC 1000714, a galaxy with a complete detached outer ring resembling Hoag's Object. Their research suggested that a recent accretion event can explain the formation of a detached outer ring. Accretion is the accumulation of particles into a massive object by gravitationally attracting more matter (usually gaseous matter) in an accretion disk [16]. A closer inspection of PGC 1000714 also revealed that its core is surrounded by two detached rings as opposed to one - an occurrence even more rare than a typical Hoag-type galaxy.

## 4.2 Convolutional Neural Networks

Artificial Neural Networks (ANNs) are biologically inspired computer programs designed to mimic the way in which the human brain processes information [6]. The ability of ANNs to detect patterns and relationships in data has led them to be known as universal function approximators. They were initially proposed in 1943, however it is modern day computation power and big data that has resulted in its success [17]. The ANN is formed with hundreds, to thousands of artificial neurons, connected by weights. They are organised in layers and can be composed with different architectures, which vary depending on the specific application. Deep neural networks (DNNs) refer to neural networks with a certain level of complexity – these networks generally contain more than two hidden layers [18]. The development of DNNs has led to state-of-the-art results in areas such as natural language processing [8] and computer vision [9].

Convolutional neural networks (CNNs) are a category of neural networks that have become the state of the art technology in areas such as image recognition and classification. They are a variation of traditional ANNs, the key difference being that a CNN uses one or more convolutional layers [19]. The use of the convolution operation in CNNs extracts features from an input image, and preserves the spatial relationship between pixels by learning image features using small blocks of input data [19].

Industry applications of CNNs can be extremely varied - from predicting human eye fixations [20], to recognition of traffic signs [21] and even the detection of bars in galaxies [22]. In the research conducted by Abraham et al., the accuracy achieved by the CNN matched the accuracy reached by human experts when presented with the same data without any additional information about the images. However, despite the capacity of CNNs to model complex objects and the ability to make mostly correct assumptions about the nature of those images, they are still seen as computationally expensive in large scale for high-resolution images [23].

## 4.3 Deep Learning in Astronomy

The use of deep neural networks (DNNs) in astronomy is an emerging topic in recent literature. Some earlier studies of machine learning for galaxy morphology classification conducted by Gauci, Adami and Abela [24] used decision tree learning algorithms and fuzzy inference systems. More recently, Kim and Brunner [25] have built effective classification systems for galaxy morphologies using DNNs. This study demonstrated over a 90% accuracy for the prediction of various aspects of galaxy morphologies using raw pixel data. Furthermore, Dieleman, Willett and Dambre [26] have also obtained such accuracies using data from the Galaxy Zoo project and specialised CNN architectures.

Further to this, in studies carried out by Tuccillo *et al*., a deep convolutional neural network was utilised to estimate galaxy luminosities and their main parameters from 2d light profile images given by optical surveys. If successful, the CNN would provide accurate morphological measurements significantly faster than other existing techniques. The results were subsequently compared to a commonly used profile fitting based software named GALFIT. Ultimately, the CNN performed equally as well as the GALFIT software, though at a factor of 500 times faster [27].

## 4.4    Search Algorithms and Optimisation Techniques

Similar to the inspiration behind ANNs, computation techniques such as Particle Swarm Optimsation (PSO) were inspired by the social behaviour of flocks of birds or schools of fish when searching for food [28]. Several studies have been conducted in order to compare these state-of-the-art heuristic optimisation techniques – Vlachogiannis and Lee (2009) experiment with numerous techniques such as directional search genetic algorithm (DSGA) evolutionary strategy optimisation (ESO) in order to solve the economic load dispatch problem in power systems [29]. The authors proposed a new variant on the PSO algorithm, an improved coordinated aggregation-based PSO (ICA-PSO), which outperformed all other benchmarked techniques for this problem. However, PSO may not always be the optimal method for a search algorithm. The choice of methods used is highly dependent on the specific use-case – for example, some methods are only effective when it is possible to compute accurate gradients of the solution with respect to variables.

# 5    METHOD

The methodology of this project begins with an exploration of the test particle simulations and the parameter space. This is followed by the selection of a search algorithm to be used to traverse the parameter space and identify any Hoag-type galaxies in the output of the simulations. A foundational component of any search algorithm is an evaluation function, described as the ability to measure the closeness between an arbitrary output and the desired output. The development of the search algorithm and the evaluation function was able to be decoupled and improved upon independently of each other. Lastly, the final procedure is outlined.

## 5.1    Test Particle Simulations

As introduced in the project scope, Dr Bekki's test particle simulations model the interaction between a gas-rich dwarf galaxy and a central elliptical galaxy. The simulations accept the following parameters:

1. The initial 3d position of the dwarf galaxy
2. The initial 3d velocity of the dwarf galaxy
3. The initial disk inclination angles ( and ) of the dwarf galaxy

Theta ( ) represents the angle between the disk of the galaxy and the − plane, and phi ( ) defines the rotation of the disk around the -axis. In this thesis, the parameter set is simplified and will comprise of:

1. The initial X position component of the dwarf galaxy
2. The initial 3d velocity of the dwarf galaxy
3. All other parameters are held constant

The simulations produce 3d particle position data that represent the state of the system over time. The number of particles used in the simulation affected the time it took for the simulation to execute. A value of 20000 particles was set in order for the search algorithm to perform with a balance between higher accuracy (with a higher number of particles) and lower execution time. Simulations of this size took approximately 30 seconds to over a minute to run, depending on the hardware specifications. Although the simulation projects the output over a time period of 12 time steps, the last time step is the focus for this research.

This data is subsequently used to create 2d images – the 3d position data is transformed to create one hundred 50x50 pixel images. Each individual image depicts the final state of the simulation from a different perspective. For our purposes, the perspective of most relevance is the face-on view – this is the image that will be used in comparison to an ideal face-on Hoag-type galaxy image. These perspectives are generated by varying two angles ( and ) from the spherical coordinates system. The angle ranges from 0 to 90 degrees in 10 consistently spaced intervals of 9 degrees, and the angle ranges from 0 to 180 degrees in 10 consistently spaced intervals of 18 degrees.

### 5.1.1    Assumptions

Based on initial investigations, the test particle simulations do not depend on any stochastic sources and use purely numerical computation to produce the output. Consistent with this, an assumption can be made that the simulations model a continuous and differentiable function. Therefore, any changes made to the input parameter set should result in a relative shift at the output. This assumption is extremely influential for deciding which search algorithm to use. For example, gradient-based or higher-order based search algorithms would not be appropriate for models that assume a discontinuous function.

## 5.2    Search Algorithm

Two methods were reviewed as possible choices for the search algorithm; a hill climber technique and particle swarm optimisation.

### 5.2.1    Hill Climbing Technique

The hill climbing technique is a mathematical optimisation technique that begins with an arbitrary solution to a problem, and aims to discover a better solution by making incremental changes to the input parameters. The performance of this technique is highly dependent on the initial starting solution. In this case, the performance is tethered to the initial parameter set for Dr. Bekki's simulations.

Hill climbing is a naïve solution that is relatively simple, which makes it a reasonable first attempt amongst optimisation techniques. However, there are two primary limitations with this method. The first drawback is that hill climbing finds optimal solutions in convex search-spaces, and in multi-dimensional polynomial problems it would only catch the local optima. We can ascertain that the simulations cannot be modelled as a convex problem, and is represented as a far more complex problem. Secondly, traditional hill climbing is not well suited for multi-dimensional search spaces. Due to these limitations, it was ultimately decided that hill climbing would not be appropriate for this problem.

### 5.2.2    Particle Swarm Optimisation

Particle swarm optimisation (PSO) is a numerical optimisation method that was introduced as an optimiser for unconstrained continuous functions [28]. The algorithm iteratively searches through a problem space as a 'swarm' of particles looking to converge on a global maximum or minimum representative of a global best solution. PSO utilises two basic equations – the position of the $n$-$th$ particle at step $(t+1)$ which is calculated by

$$x_{t+1} = x_t + v_t$$    (1)

where $v(t)$ is the particle's velocity. The velocity of the particle at step $(t+1)$ is given by

$$v_{t+1} = \omega v_t + c_1 r_1 (p_t - x_t) + c_2 r_2 (g_t - x_t).$$    (2)

In these equations, values $c_1$ and $c_2$ are the acceleration coefficients. $c_1$ is the cognitive coefficient and influences particles to return to individual best regions of the search space. $c_2$ is the social coefficient, and directs particles to move to the best region the swarm has found (i.e. global best solution). $\omega$ is the inertial coefficient which affects the trade-off between convergence and exploration during the search process. Each particle maintains their position in the search space, their velocity    and individual best position    . The summation of the particles, known as the swarm, maintains the global best position    . The non-deterministic nature of the algorithm stems from the two random numbers, $1$ and $2$, that are present in the particle velocity formula.

The selection of PSO parameters have a significant impact on the optimisation performance [39]. Substantial research has been made into this area, and general guidelines do exist. These guidelines have been adopted by researchers also applying PSO to optimisation and search problems. Therefore, the Standard PSO 2006 [30] coefficient values were selected.

**Table 2:** The PSO coefficient values selected for this research based on the Standard PSO 2006 values.

| Parameter | Value | Description |
|---|---|---|
| $w$ | $12\log(2)=0.72$ | Inertia coefficient |
| $1$ | $0.5+\log 2=1.19$ | Cognitive coefficient |
| $2$ | $0.5+\log 2=1.19$ | Social coefficient |
| $n$ | 40 | Number of particles |

Upon random initialisation of particle positions, the fitness of each particle is calculated by an evaluation function. Following this, the individual and global best solutions are updated. This information is then used to calculate the position and velocity of each particle. This process is continuously repeated until some termination criteria is met. The termination condition set the algorithm to stop after a maximum number of iterations have been completed.

## 5.3    Evaluation Function

The success of the search algorithm relies heavily on the accuracy and correctness of the evaluation function used. Initially, a common similarity metric – cosine similarity – was trialled as the evaluation function by comparing the image of the face-on perspective from the simulation output to an image of an ideal Hoag-type galaxy from its face-on view. Due to the limitations with this method, a two-fold method was conceived using a convolutional neural network (CNN) as a filter for ring galaxies (Hoag-type galaxies are a subset of ring galaxies) before applying cosine similarity. The next iteration of the evaluation function proposed the use of cosine similarity on feature vectors extracted by passing the simulation output and ideal Hoag-type output through the CNN. Further research and experimentation then led the evaluation function to consist primarily of a neural network architecture known as a Siamese network.

### 5.3.1    Cosine Similarity

Traditional image comparison techniques such as the histogram method can be relatively susceptible to variations in noise, rotation and scale. Content based retrieval systems and other systems that require image similarity have attempted to transform a pair of images to their respective vectors and apply a distance function such as cosine similarity between them [37]. Cosine similarity is mathematically defined as

$$\cos\ =\quad =\quad \cdot \tag{3}$$

where ⋅ is the dot product of the vectors and ; and being the magnitude of the vectors. This approach falls under the category of pixel by pixel comparisons, which are known to not be robust to noise variations and small variations to position and rotation. It was concluded that cosine similarity was not a suitable means as an evaluation function, so further research was conducted.

### 5.3.2 Convolutional Neural Network and Cosine Similarity

A convolutional neural network (CNN) was successfully trained to classify disk, shell and ring galaxies to act as a filter for the simulation output. If the face-on perspective of the output image was classified as a ring galaxy by the CNN, cosine similarity was subsequently applied to the output image and the ideal Hoag-type galaxy. This pipeline was created with the intention of improving on the initial naïve solution of using cosine similarity only.

**Dataset Creation**

Dr. Bekki provided additional software to generate different types of galaxy images to train the network on. This software produced images of disk, shell, ring and Hoag-type galaxies. Several parameters could be configured to change the properties of the chosen galaxy type. For example, in order for the simulation to output Hoag-type galaxies, the inner-edge of a ring galaxy was required to be set between certain values. Figure 3 displays some of the possible outputs from the software.

**Figure 3**:

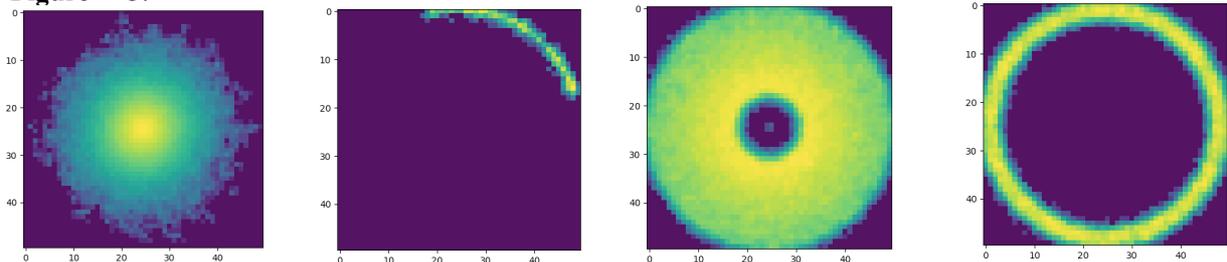

Disk       Shell       Ring       Hoag-type

Images of disk, shell ring and Hoag-type galaxies generated by Dr. Bekki's software. These particular images represent the face-on perspective of the galaxy.

60,000 images were generated and collected to form the dataset for the CNN. As neural networks can be vulnerable to bias in the training data, it was necessary to ensure that the training data would contain an even distribution across the possible classes. There were 15,000 images of disk, shell and ring galaxies as a result of 150 different configurations for each type of galaxy. A further 15,000 images were empty and represented no galaxy formation. This dataset was shuffled and split so that 45,000 images made up the training data, and the remaining 15,000 images were arranged as the testing data in a 75/25% split.

**Training Process**

The training process begins by initialising all neuron weights randomly. During the forward pass of the network, the input images are fed to the network and the loss is calculated. The error is then back-propagated to update the weights using gradient descent during the backwards pass.

Backpropagation is a crucial method in the success of training neural networks. As part of this algorithm, the derivative of the loss function with respect to the network is required.

While there are several different kinds of loss functions, cross-entropy loss was chosen due to its suitability for classification tasks. The cross-entropy loss function increases as the predicted probability diverges from the true probability, and decreases as the predicted probability converges to the true probability [31]. Cross-entropy loss is defined as

$$= - \sum_{=1} ' \log \qquad (4)$$

where $'$ is the predicted probability for class  from the network and  is the true probability for class  .  represents the number of class in the network.

These processes repeat until the loss of the network converges or the training process is manually interrupted. The input images can be fed to the network singularly, or in batches. While larger batch sizes can speed up the time to convergence, they also subsequently have higher memory requirements. Selecting batch sizes amongst other hyper-parameters such as learning rates, activation functions etc. fall under the convoluted branch of neural network design.

**Hyper-parameters**

In practice, the design of a neural network model is considered to be a hyper-parameter optimisation problem [32]. There are no strict guidelines as to how to design a neural network – for example, the number of hidden layers or size of the convolutional filters.

Numerous neural networks architectures were tried and tested by varying multiple hyper-parameters. However, not all combinations could be explored due to time constraints and the eventual discovery of an architecture that achieved an extremely high accuracy. The hyper-parameters that were investigated were:
1. The number of kernels per convolutional layer (8, 16, 32, 64, 128)
2. The size of the kernel in each layer (3, 4, 5, 6, 7, 8, 9)
3. The number of convolutional layers (2, 3, 4)
4. The number of max-pooling layers (1,2,3)
5. The number of fully connected layers (1, 2)

The accuracy of the network on testing data was used as the quantifiable metric to compare performance of different architectures. The approach for adjusting the hyper-parameters consisted of trial and error, with some domain knowledge guiding these decisions. This process of iteratively improving the neural network architecture is consistent with current research and literature [40]. Other noteworthy hyper-parameters included the learning rate, which was set to 0.001 and the training batch size, which was set to 100.

**Implementation**

The CNN models were created using the software library TensorFlow in Python. TensorFlow is an open-source library that is commonly used for machine learning applications such as neural networks. It enables users to define, train and deploy machine learning models, and has been optimised for numerical computation.

While this combination of using a trained CNN acting as a filter and subsequently applying cosine similarity to ring galaxies was an improvement, it still fundamentally worked as a pixel by pixel comparison method. This led to the progression of using the extracted feature vectors of an image that was passed to the trained CNN model. It was more feasible to use the CNN as a way of finding the most important regions and identifiers of an image and to then apply the distance function to compare these extracted regions.

### 5.3.3 Siamese Network

Siamese networks are a natural evolution from the idea of applying cosine similarity to the feature vectors extracted from the CNN when two images are to be compared. They are a particular type of neural network architecture – instead of a model that learns classification tasks, the model learns to differentiate between two distinct inputs. This can be conceptualised as learning the similarity between two images. Siamese networks are constructed in such a way that twin neural networks each accept distinct inputs which are then passed to a shared loss function. The architecture and hyper-parameters of the twin neural networks are identical and they utilise a technique known as weight sharing to ensure two similar images are mapped by each network to the same regions in feature space [33]. In this case, the twin networks are CNNs that are each fed an image, and a loss function is then used to compute a metric between the outputs of the networks.

**Dataset Creation**

A smaller dataset of images was used to train the Siamese network. Sharing weights across the twin networks means fewer parameters to train for, which means less data was required. 2,000 images of each class – disk, ring, shell and Hoag-type galaxies – were generated for the training data. An additional 100 images of each class was created to make up the testing data.

**Training Process**

The training process of the Siamese network is quite similar to the CNN training process; each image from a pair of images is fed into a CNN and the network weights are trained by minimising the loss. However, the conventional cross-entropy loss function is replaced by the contrastive loss function. In contrast to conventional learning systems where the loss function is a sum over samples, the contrastive loss function runs over pairs of samples. The contrastive loss function was presented by LeCun *et al.,* and works to decrease the loss of similar pairs and increase the loss of dissimilar pairs [34]. The function is defined as

$$= 1 - 12 \quad 2 + ( )12 \max(0, - )2 \qquad (5)$$

where    is the Euclidean distance between the outputs of the twin Siamese networks. The Euclidean distance is represented mathematically as

$$= \quad 1 - 22 \qquad (6)$$

with    defined as the output of one of the twin networks and   *1*,  *2* as the pair of images input to the system [34]. The value    is either equal to 0 or 1, depending on if the inputs are from the same class. For example, if the inputs are both disk galaxies then    will be 1. The value    is the margin value and must be greater than 0. This margin establishes a cut-off value so that dissimilar pairs with a loss higher than the margin won't contribute to the overall loss of the network. This is to ensure the network is optimised on image pairs that the network believes are similar, but are actually dissimilar. The    function merely chooses the highest value from the pair of numbers it is provided.

The Siamese network is also trained until the loss of the network converges, or the training process is manually interrupted. As the twin networks of the Siamese network architecture were set to be CNNs, the relevant hyper-parameters were noted as the same as *5.3.2 Hyper-parameters.*

**Implementation**

The Siamese model was implemented using PyTorch, an open source neural network library for Python. PyTorch is relatively new in comparison to a well established library like TensorFlow, however it is quickly gaining recognition for its ease-of-use and lower learning curve. While TensorFlow conceptualises its machine learning models as static graphs, PyTorch uses dynamic graphs. This difference in abstraction has made PyTorch relatively easier to work with to code and build a deep learning model and is the reason it was selected as the tool to create the Siamese network.

**5.4    Final Procedure**

The use of the trained Siamese network as the evaluation function $f( )$ – where is a particle's position in the search space – of the PSO algorithm has not been found in existing literature and research. Thus it may be postulated that this could be a novel combination discovered through the process of this thesis. The final procedure consisted of using this two-fold method to traverse the simulation parameter search-space. The PSO particles' positions at one time-step represented a set of parameters for the simulation, and in order to evaluate the correctness of each parameter set, the Siamese network was used to determine the similarity between the parameter set and an ideal Hoag-type galaxy. The goal of the particles was to *minimise* this dissimilarity value, essentially find the global minima. The search algorithm was set to run for a maximum number of iterations, and the parameter set that resulted in the closest similarity to a Hoag-type galaxy was output.

# 6    RESULTS AND DISCUSSION

## 6.1    CNN Performance

Table 3 below summarises the different CNN architectures that were tested alongside their respective performances.

**Table 3:** Summary of number of attempts, architecture of each attempt, and performance of each attempt based on the accuracy of the model on testing data.

| Attempt | Architecture | Performance (Accuracy on testing data) |
|---:|---|---|
| 1 | 2 Convolution layers<br>2 Max-pool layers<br>1 Fully-connected layer | 95.1% |
| 2 | 2 Convolution layers<br>2 Max-pool layers<br>2 Fully-connected layer | 98.4% |
| 3 | 3 Convolution layers<br>2 Max-pool layers<br>1 Fully-connected layer | 96.8% |
| 4 | 3 Convolution layers<br>3 Max-pool layers<br>2 Fully-connected layer | 99.6% |

The best performing model achieved an accuracy of 99.6% and consists of 3 convolutional layers with 3 sets of max pooling layers - the max pooling layers perform down sampling of an input layer in a non-linear fashion to reduce the computational complexity. Each convolutional layer is succeeded by a Rectified Linear Unit (ReLU) [35], a type of activation layer. This is followed by two fully connected layers. The first fully connected layer drops neurons that have weak connections whose weights are not sensitive to weight updates. This operation is known as dropout, and is a method of avoiding model overfitting. The dropout in this model is 40%. The final fully connected layer calculates the cross-entropy loss during training, and outputs a soft-max probability score for each possible class. The model trained for 10,000 iterations before converging to a loss of 0.01097. Figure 4 below illustrates this model architecture.

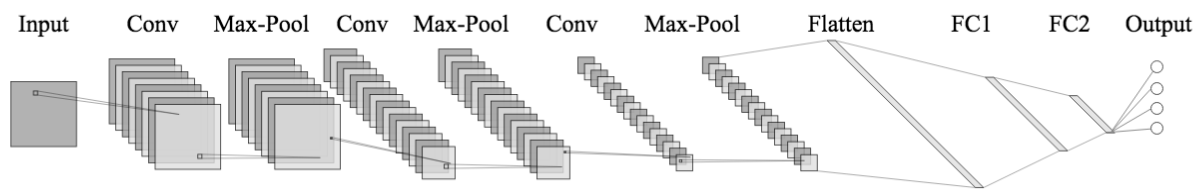

**Figure 4**: Architecture diagram of the best performing CNN model.

## 6.2    Siamese Network

The architecture of the twin convolutional neural networks in the Siamese network is summarised in the table below. A slight variant of the best performing CNN model was used for the Siamese network, as it was already proven to perform well on the dataset. One of the fully connected layers was removed in order to not overtly down-sample the final vector which would be passed to the Euclidean distance function. Since the hyper-parameters and architecture of each twin network is symmetrical, Table 4 reflects the model used for both networks and Figure 5 defines the architecture of the overall Siamese network.

**Table 4:** Definition of each layer in the twin networks.

| Layer | Input | Weights | Details |
|---|---|---|---|
| Convolution | 50x50 | 32x5x5 | ReLU activation |
| Max-Pool | 32x46x46 | - | 2x2 kernels |
| Convolution | 32x23x23 | 64x5x5 | ReLU activation |
| Max-Pool | 64x20x20 | - | 2x2 kernels |
| Convolution | 64x10x10 | 64x3x3 | ReLU activation |
| Max-Pool | 64x8x8 | - | 2x2 kernels |
| Flatten | 64x6x6 | - | - |
| Fully Connected 1 | 2304 | 2304x128 | ReLU activation |

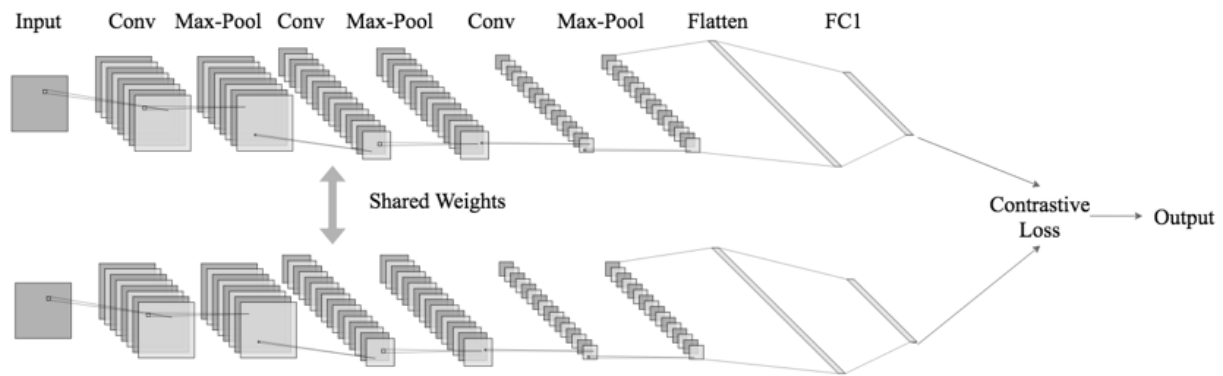

**Figure 5**: Architecture diagram of the Siamese network model.

The final model of the Siamese network took approximately 6,000 iterations to converge as can be seen in Figure 6. The graph illustrates the contrastive loss on the y-axis and the iteration count on the x-axis, and it is observed that the contrastive loss drops quite early during the training process.

There are no formal methods for evaluating the performance of the Siamese network, unlike traditional CNNs which are usually assessed based on the accuracy of the model on testing data. To calculate the similarity between two images, the images are first passed through the Siamese network. The output vectors are then used to calculate the Euclidean distance *(   )* which directly corresponds to the dissimilarity between the image pair. The higher the value of    , the higher the dissimilarity. To assess the quality of the network, the dissimilarity is calculated for all combinations of a Hoag-type image and an image from one of the other galaxy classes. This is repeated for each of the other classes using the 100 images of each class in the testing dataset and the average is calculated from the 100x100 combinations per class. This is illustrated in Figure 7. A relatively low average dissimilarity is found between the Hoag-type and ring galaxies. Conversely, a higher average is found when calculating the dissimilarity of the disk and shell galaxies to the Hoag-type galaxies. These results indicate

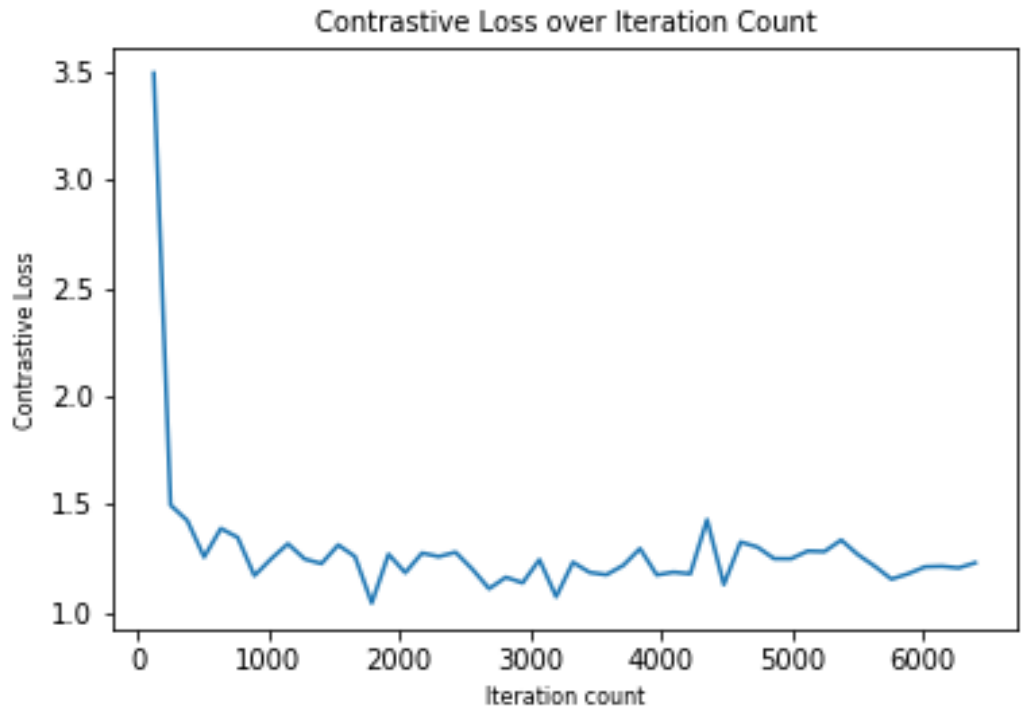

the Siamese network is functioning as expected.

**Figure 6**: The contrastive loss over the number of iterations the Siamese network was trained for.

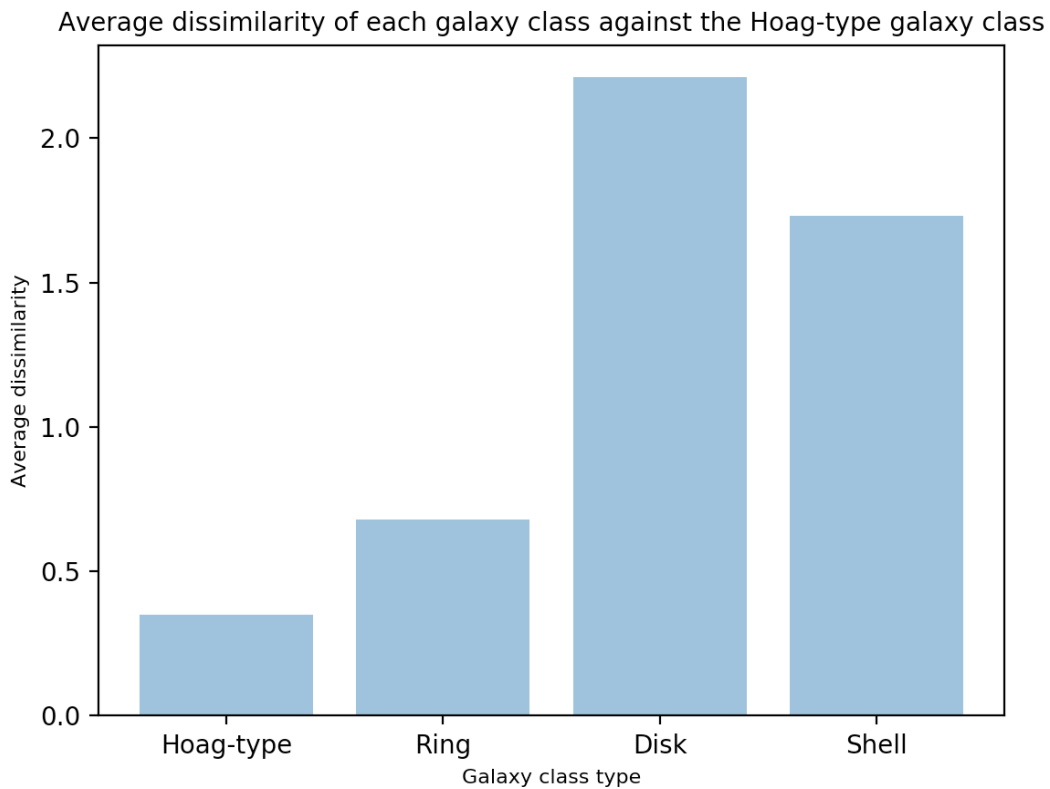

**Figure 7**: The average dissimilarities of the galaxy classes against the Hoag-type galaxy class.

## 6.3    Best Parameter Set

This parameter set was achieved with 40 particles which terminated after 69 iterations. An interesting area to analyse is how the PSO particles move over time and converge to a solution. As a 4-dimensional search-space can be difficult to visualise, Figure 8 below shows a 3d projection of one simulation's worth of explored parameter sets. This is done by ignoring the X-position parameter and using a colour to represent the dissimilarity. Each point represents one parameter set; they begin dispersed in the parameter space as they are randomly initialised. Over time, two (blue) clusters appear that indicate the solutions are close matches to Hoag-type galaxies. It can be deduced that these two areas represent the local minima of the parameter space.

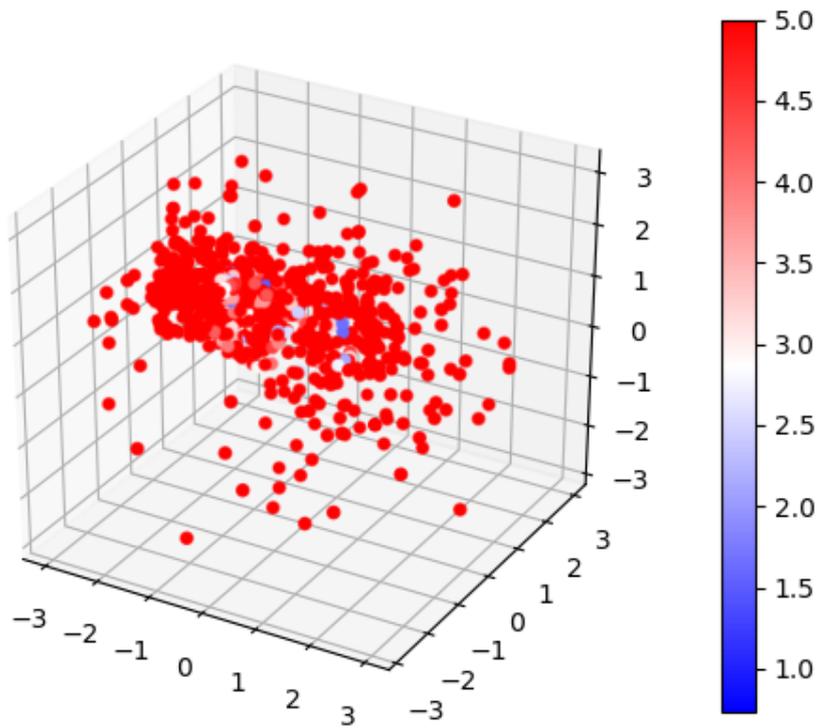

**Figure 8**: A 3d projection of the Y velocity parameter sets explored during one simulation.

The best parameter set that resulted in the closest match to the ideal Hoag-type galaxy is outlined in Table 5.

**Table 5:** Summary of the best parameter set resulting in the closest match to the ideal Hoag-type galaxy.

| Parameter | Value | Boundaries | Description |
|-----------|-------|------------|-------------|
| X-position | 6.014 | (5,10) | X position of dwarf galaxy |
| X-velocity | -0.150 | (-3,3) | X velocity of dwarf galaxy |
| Y-velocity | 1.426 | (-3,3) | Y velocity of dwarf galaxy |
| Z-velocity | 0.024 | (-3,3) | Z velocity of dwarf |

galaxy

The face-on view image of this parameter set can be seen in Figure 9 below.

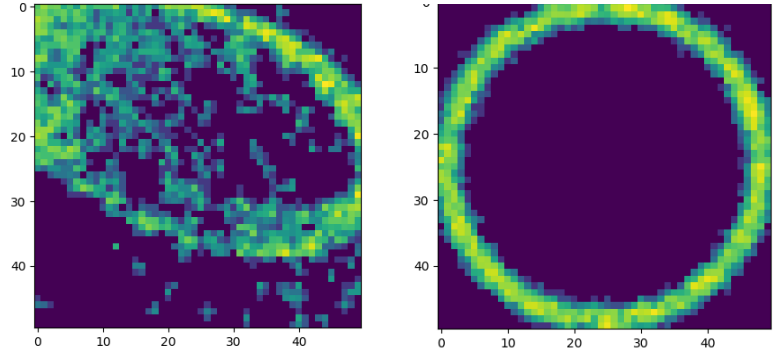

**Figure 9**: (Left) Face-on view of the best parameter set. (Right) Face-on view of the ideal Hoag-type galaxy.

Figure 9 displays the face-on view of the best parameter set compared to the ideal Hoag-type galaxy. The final dissimilarity value was 0.83.

## 6.4    Validation of Best Parameter Set

It was necessary to validate this result using more complex and accurate simulations written by Dr. Bekki. The test particle simulations that the search algorithm ran on only took into consideration the effects of gravitational dynamics and thus represent a simplified model. However, Dr. Bekki has written chemodynamical simulations from 2013 [36] that additionally accounts for hydrodynamics, chemical evolution and dust formation and evolution. Hydrodynamics were important to consider as they have been cited as fundamental to certain processes such as galaxy formation [36]. Each individual execution of the more complex simulation could take between 24 to 48 hours to complete. This means evaluating one parameter set can take a very long time, hence the reasoning as to why the search algorithm is run on test particle simulations (~30 seconds to a minute per run). By using the parameter set in Table 5 as an initial estimate, Dr. Bekki was able to produce a fairly close match to a Hoag-type galaxy. This result can be seen in Figure 10, which showcases 6 images over time, with the last image depicting a galaxy with a distinct ring matching that of a Hoag-type galaxy.

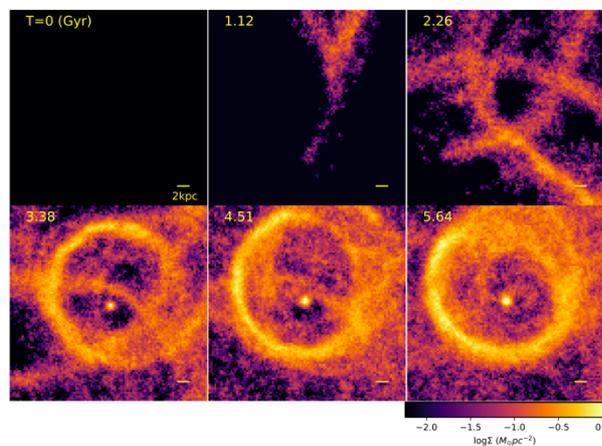

**Figure 10**: Output from full chemodynamical simulations using the best parameter set in Table 5 as an initial estimate.

## 6.5    Repeated Simulations

The search algorithm was run multiple times and converged to different parameter sets. This is to be expected due to the stochastic and non-deterministic nature of the PSO algorithm, which will not guarantee the same result each time the search is executed. Table 6 below showcases other results that were achieved where the output had resembled an ideal Hoag-type galaxy.

**Table 6:** Additional parameter sets output from the search algorithm that beared resemblance to the Hoag-type galaxy.

| | Parameter Set | Dissimilarity | Output Image (Face-On) |
|---|---|---|---|
| *1* | X-position = 6.545<br>X,Y,Z-velocity = 0.475, 0.591, -0.022 | 1.02 | 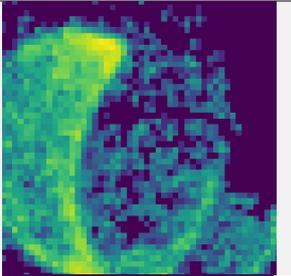 |
| *2* | X-position = 5.000<br>X,Y,Z-velocity = -0.250, 1.444, 0.839 | 0.81 | 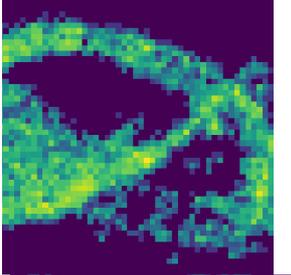 |
| *4* | X-position = 5.000<br>X,Y,Z-velocity = -1.651, 1.043, -0.812 | 0.93 | 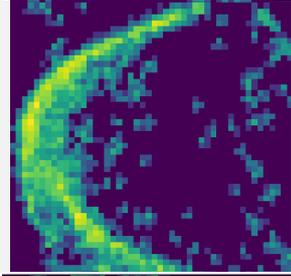 |
| *5* | X-position = 8.708<br>X,Y,Z-velocity = -1.574, 1.217, -0.566 | 0.87 | 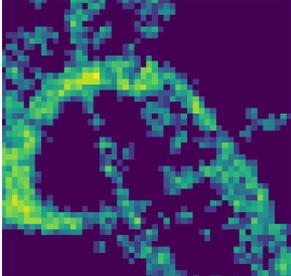 |

Due to time constraints, not all of these parameter sets could be validated against the gravitational dynamics and hydrodynamic simulations.

## 6.6    Local Minima

One potential limitation of the solution is that the PSO algorithm particles are being caught in local minima. Unfortunately, there are few search algorithms that can guarantee that the global optimum solution is found in complex, multidimensional polynomial search spaces. Brute forcing may be a possible solution, as every conceivable parameter set would be explored. However, this is rarely a practical solution and in this problem, it would take an almost infinitely long time for a brute force method to execute. If the particles are constantly being trapped into local optima, the algorithm is not exploring enough of the search space and the parameters should be changed in an attempt to reverse this behaviour.

As discussed in *5.2.2 Particle Swarm Optimisation*, it is required to set a social coefficient $(c_2)$ that influences the likelihood of a particle following the global best solution. If this value is set too high, it is probable that the particles are too heavily influenced and restricting the search space. The cognitive coefficient $(c_1)$ and the inertia coefficient $(w)$ may also be set too low, as these coefficients influence particles to return to individual best regions of the search space and affects the trade-off between convergence and exploration during the search process respectively. With additional time, it would have been possible to further explore the effects of the PSO parameters on the results of the problem being analysed. Another simple method for overcoming the local minima could be to simply run the algorithm for a longer period of time – it could have been prematurely terminated. Thinking beyond the scope of this project, alternative search algorithms and optimisation techniques could also be explored.

## 6.7    Over-fitting of Siamese Network

One problem that can manifest with neural networks is over-fitting. This occurs when the error of the training set has converged to a small value, but is much larger when new data is passed through the network. Additionally, neural networks that have sufficient parameters relative to the size of the training data may be deemed over-fitted. Over-fitted neural networks perform extremely well on the training data but are not able to generalise to new and unfamiliar data. The Siamese network that was trained does contain mechanisms to prevent this from happening, applying a dropout layer which is commonly used to prevent over-fitting of a network.

It is also possible that the Siamese network does not perform optimally over the output of the simulations. The images that made up the training dataset are much less noisy than some of the types of images output by the simulations. Whilst this did not affect the performance of the CNN model trained to classify the galaxy types that achieved an accuracy of 99.6% on testing data, it may have potentially impacted the results of the Siamese network which has not been exposed to noisier images. It would also be feasible to re-train the Siamese network on a new dataset, and manually add noise to images to increase the robustness of the model.

## 6.8    Hidden Layer Analysis

Neural networks are often referred to as 'black boxes' due to the lack of comprehensive understanding of how these models work, and in particular, what computations they perform at intermediate layers. One suggestion on how to deepen this knowledge involves visualising and interpreting the activations produced on the layers of a trained neural network. Observing these activations assists in building important intuitions about how convolutional networks behave. The kernels from the first convolutional layer of the Siamese network are shown below.

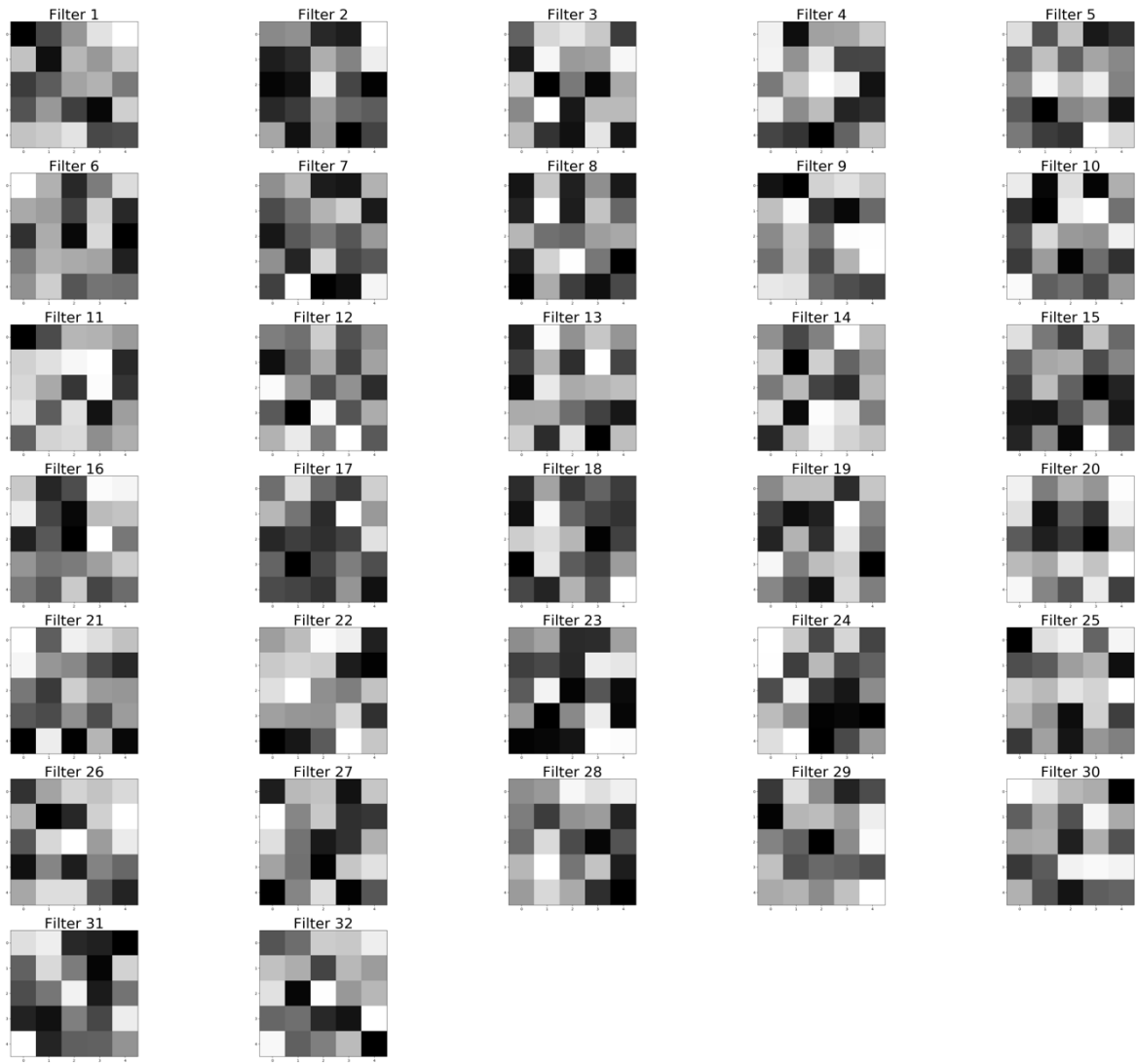

**Figure 11**: Extracted kernels from the first convolutional layer from the Siamese network.

These convolutional kernels alone are difficult to interpret and may appear meaningless to the human eye. However, applying the kernels to a regular image provides interesting insight as to what features and information the convolutional layer is extracting. The kernels are applied to Figure 12, a black and white image of a cat.

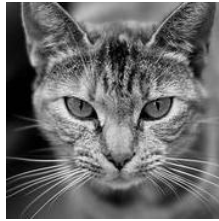

**Figure 12**: A black and white image of a cat.

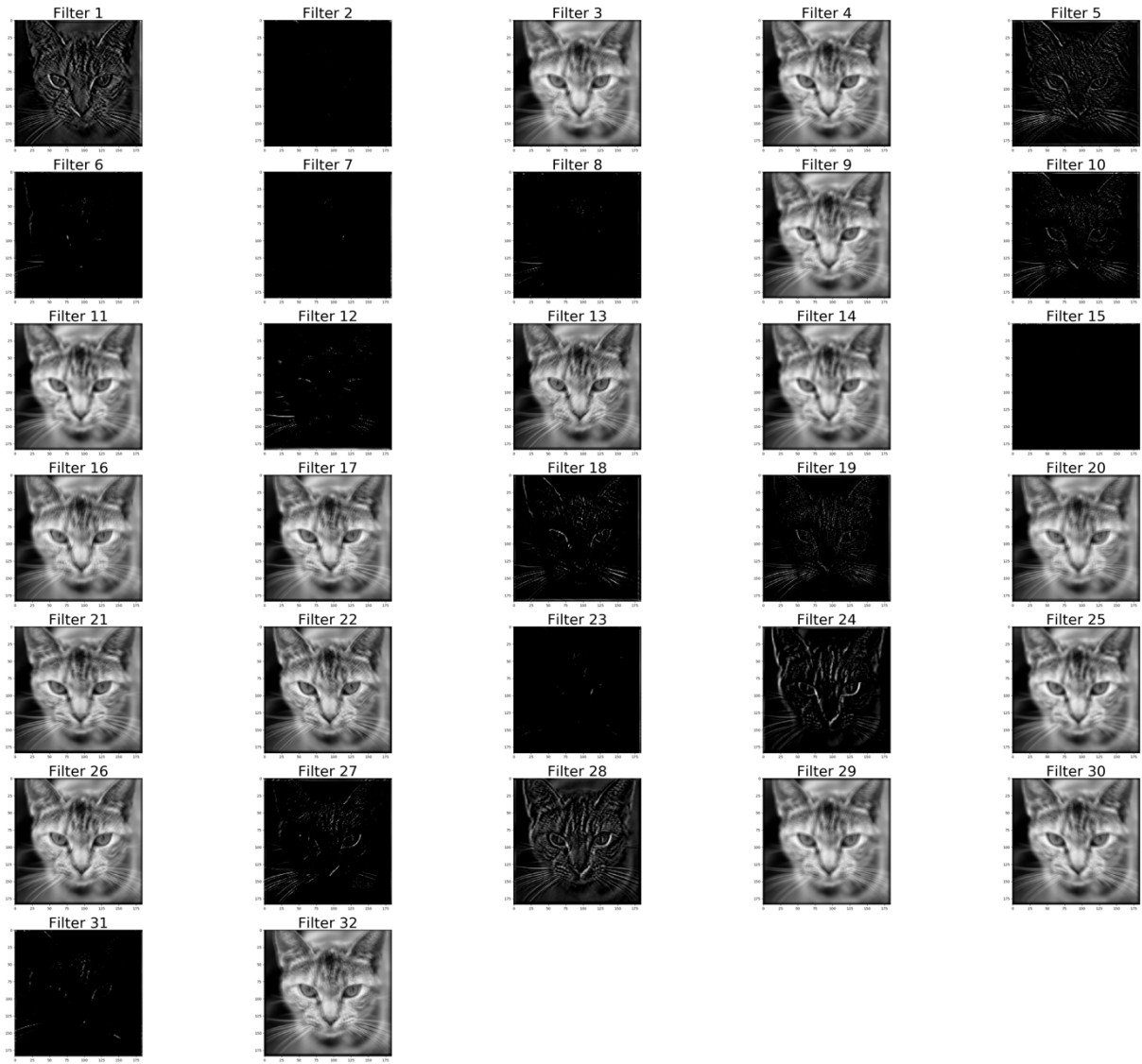

**Figure 13**: The first convolutional layer kernels applied to the image of the cat in Figure 12.

Each kernel learns to activate optimally for different features of the image. Although there appear to be some kernels that have little to no effect, there are kernels that seem to be acting as edge detection kernels. Filters 1, 5, 19, 24, 27 and 28 in Figure 13 exhibit some degree of this behaviour with varying intensities. Edge detection aims to identify regions in an image where the image brightness abruptly changes or experiences discontinuities. The different edge detecting kernels also appear to emphasise various regions or types of edges. These observations suggest that the network is extracting the gradient information throughout the process. However, there are a number of filters that do not appear to have any effects. This could indicate that the network is over-fitted and is not using all the filters available.

## 6.9    Implications

This result is significant as it confirms that this project was successful in creating a way to decrease the time taken to explore the simulation parameter space. With the increase in telescope technology, the amount of information gathered about stellar objects has increased dramatically [38] to the point where manual investigation of this data may no longer be the best option. Researchers want to be able to increasingly make use of software and automation to complete these time-consuming tasks. For example, the Galaxy Zoo project conducted by Buta involved the manual categorisation of 3962 ringed galaxies [1]. However, neural networks have the potential to complete this work in a much faster timeframe.

# 7    CONCLUSION

The motivation behind this research was to firstly, study and model the formation process of Hoag-type galaxies as a simulation and secondly, dramatically decrease the time it would take to search the simulation parameter space. This led to an investigation with the aim – to determine the parameter set for Dr Bekki's test particle simulations that would most closely result in an output resembling a Hoag-type galaxy. This required the creation of a search algorithm to be used in the parameter search-space, as well as an evaluation function that is necessary to the success of the search algorithm.

From a naïve evaluation function using cosine similarity on the image pixels, the evaluation function evolved to use convolutional neural networks and eventually Siamese networks. The optimal CNN model architecture was discovered through an educated process of trial and error, which is accepted as the de facto standard in industry and research. Due to the high level of accuracy achieved by the CNN on galaxy classification tasks – 99.6% – on testing data, this model was utilised as the architecture of each respective twin network in the Siamese network. Siamese networks were then able to learn the dissimilarity between Hoag-type galaxies and other galaxy classes.

The combination of using a trained Siamese network as the evaluation function for the PSO algorithm has not been found in existing literature to date. This two-fold method was able to successfully find a parameter set for the simulations that resembled a Hoag-type galaxy. Additionally, the simulation process could also be considered a two-fold procedure; wherein the results from the test particle simulations were validated against the full chemodynamical simulations, which require large amounts of computing time (ranging from 24 to 48 hours per run). A good match to a Hoag-type galaxy was found in the chemodynamical simulations, supporting the hypothesis that Hoag-type galaxies are formed through the interaction of a gas-rich dwarf galaxy and central elliptical galaxy. Following the proven results from the initial hypothesis and subsequent investigation, it is proposed that this novel two-fold method be further developed and used in other areas.

# 8 FUTURE DIRECTIONS

## 8.1 Larger Parameter Space

The search algorithm utilised is currently limited to only 4 dimensions (X position and X, Y, Z velocity) of the parameter space. However, there are 6 dimensions in total including the Y and Z position components and many other parameters that affect the results of the simulation. This work should logically be extended and tested in the full parameter space to further assess the effectiveness of the solution. As a result, the search algorithm will be able to traverse and explore this new parameter space. The effect of holding these parameters constant could have had a considerable impact on the best solution that was found, as they represent key pieces of information within the simulations.

## 8.2 Alternative Search Algorithms

This research could be possibly extended by utilising different optimisation algorithms. Only 2 algorithms – hill climbing technique and particle swarm optimisation – were investigated and trialled. There are a plethora of search algorithms and optimisation methods that may perform well in this particular search-space. In particular, deep reinforcement learning is a potential candidate solution as it is currently at the forefront of machine learning research. As reinforcement learning encompasses goal-oriented algorithms that learn how to accomplish a complex objective; it is far more complicated than using PSO. Therefore, it may potentially yield more improved and interesting results.

## 8.3 Other Applications of Novel Technique

Future work on this project would benefit from exploring and performing additional analysis on the novel combination of PSO and Siamese networks as the evaluation function. This technique could potentially be applied to any search problem where the desired output is known beforehand. There are potential use cases within astronomy and outside in other industries which may be suitable, and it would be of great interest to compare the performance between conventional evaluation functions and these novel methods in the future.

# 9    ACKNOWLEDGEMENTS

Thank you to Dr. Kenji Bekki and Dr. Lyndon While for their guidance and assistance throughout the project.